\documentclass[a4paper,12pt]{article}
\usepackage{graphicx,rotating,colortbl,xcolor,wrapfig,hyperref,slashed,cite}

\hypersetup{colorlinks,bookmarksopen,bookmarksnumbered,
linkcolor=blus,pdfstartview=FitH,urlcolor=rossos,citecolor=verde}

\def\to{\rightarrow}
\def\bea{\begin{eqnarray}}
\def\eea{\end{eqnarray}}

\def\meq#1{~(\ref{eq:#1})}
\def\mpl{{\bar{M}_{\rm Pl}}}

\newcommand{\sfrac}[2]{{#1/#2}}

\definecolor{Gray}{gray}{0.95}
\newcommand{\bbox}[1]{\fcolorbox{gray}{Gray}{~$\displaystyle #1$~}}

\newcommand{\lx}{\lambda}

\newcommand{\Rc}{{\cal R}}

\definecolor{rosso}{cmyk}{0,1,1,0.4}
\definecolor{rossos}{cmyk}{0,1,1,0.55}
\definecolor{rossoc}{cmyk}{0,1,1,0.2}
\definecolor{blu}{cmyk}{1,1,0,0.3}
\definecolor{blus}{cmyk}{1,1,0,0.6}
\definecolor{bluc}{cmyk}{1,1,0,0.1}
\definecolor{verde}{cmyk}{0.92,0,0.59,0.25}
\definecolor{verdec}{cmyk}{0.92,0,0.59,0.15}
\definecolor{verdes}{cmyk}{0.92,0,0.59,0.4}

\newcommand{\Sb}{\mathcal{S}}

 \def\be   {\begin{equation}}   \def\ee   {\end{equation}}
 \def\ba   {\begin{array}}      \def\ea   {\end{array}}

\font\tenrsfs=rsfs10 at 12pt
\font\sevenrsfs=rsfs7
\font\fiversfs=rsfs5
\newfam\rsfsfam
\textfont\rsfsfam=\tenrsfs
\scriptfont\rsfsfam=\sevenrsfs
\scriptscriptfont\rsfsfam=\fiversfs
\def\mathscr#1{{\fam\rsfsfam\relax#1}}

\newcommand{\bp}{\bar M_{\rm Pl}}

\newcommand{\GeV}{\,{\rm GeV}}

\def\circa#1{\,\raise.3ex\hbox{$#1$\kern-.75em\lower1ex\hbox{$\sim$}}\,}

\newcommand{\beq}{\begin{equation}}
\newcommand{\eeq}{\end{equation}}
\newcommand{\hmax}{h_{\rm max}}

\font\ital=cmu10 

\makeatletter
\def\hhref#1{\href{http://arxiv.org/abs/#1}{arXiv:#1}}
\usepackage{xstring} 
\newcommand{\hhrefq}[1]{\IfSubStr{#1}{:}{\href{http://inspirehep.net/search?ln=en&ln=en&p=#1&of=hb&action_search=Search&sf=&so=d&rm=&rg=25&sc=0}{InSpires:#1}}{\hhref{#1}}}

\def\art{\@ifnextchar[{\eart}{\oart}}
\def\eart[#1]#2#3#4#5#6{{\rm #2}, {\em #3 \bf #4} {\rm (#6) #5} ({\em #1})}
\def\article{\@ifnextchar[{\earticle}{\oarticle}}
\def\oarticle#1#2#3#4#5#6{{\rm #1}, {\ital ``#6''}, {\rm #2 #3 (#5) #4}}
\def\earticle[#1]#2#3#4#5#6#7{{\rm #2}, {\ital ``#7''}, {\rm #3 #4 (#6) #5}  [\hhrefq{#1}]}
\def\hepart[#1]#2{{\rm #2, \ital#1}}
\def\heparticle[#1]#2#3{#2, {\ital ``#3''} [\hhrefq{#1}]}
\newcommand{\doi}[1]{\href{http://dx.doi.org/#1}{[link]}}

\def\hhref#1{\href{http://arxiv.org/abs/#1}{arXiv:#1}} 

 \def\lx{\lambda}

\def \lta {\mathrel{\vcenter
     {\hbox{$<$}\nointerlineskip\hbox{$\sim$}}}}
\def \gta {\mathrel{\vcenter
     {\hbox{$>$}\nointerlineskip\hbox{$\sim$}}}}

\begin{document}
CERN-TH-2017-121   \hfill IFUP-TH/2017

\thispagestyle{empty}
\vspace{0.1cm}
\begin{center}
{\huge \bf \color{rossos} (Higgs)
vacuum decay\\[2mm] during inflation}  \\[2cm]

{\bf\large Aris Joti$^a$, Aris Katsis$^a$, Dimitris Loupas$^a$, Alberto Salvio$^b$, Alessandro Strumia$^{b,c}$, Nikolaos Tetradis$^a$, Alfredo Urbano$^b$}  \\[5mm]

{\it $^a$ Department of Physics, National and Kapodistrian University of Athens, Greece}\\[1mm]
{\it $^b$ } {\em CERN, Theoretical Physics Department, Geneva, Switzerland}\\[1mm]
{\it $^c$ Dipartimento di Fisica dell'Universit{\`a} di Pisa and INFN, Italia}

\vspace{2cm}
{\large\bf\color{blus} Abstract}

\begin{quote}\large
We develop the formalism for computing gravitational corrections to
vacuum decay from de Sitter space
as a sub-Planckian perturbative expansion.
Non-minimal coupling to gravity can be encoded in an effective potential.
The Coleman bounce continuously deforms into
the Hawking-Moss bounce, until they coincide for a critical value of the Hubble constant. 
As an application, we reconsider the decay of the electroweak Higgs vacuum during inflation.
Our vacuum decay computation reproduces and improves bounds on the maximal inflationary Hubble scale
previously computed through statistical techniques.

\end{quote}
\end{center}

\newpage
\tableofcontents
\normalsize


\section{Introduction}
A false vacuum can decay through quantum tunnelling that leads to nucleation of regions of true vacuum.
The rate of this non-perturbative phenomenon is
exponentially suppressed by the action of the `bounce', 
the field configuration that dominates the transition \cite{Coleman:1977py}. 

Coleman and de Luccia showed how to account for gravitational effects~\cite{deluccia}.  
Their formalism can be simplified by restricting to sub-Planckian energies, which are the only ones for which
Einstein gravity can be trusted.
Simplified expressions were obtained in~\cite{rychkov,urbano} for the decay of flat space-time.

\smallskip

We extend here the simplified formalism to tunnelling from de Sitter space (positive energy density), 
which is relevant during inflation with Hubble constant $H$.
It is known that gravitational effects can dramatically enhance the tunnelling rate.
The qualitative intuition is that a de Sitter space has a Gibbons-Hawking 
`temperature' $T=H/2\pi$~\cite{gibbons}
that gives extra `thermal' fluctuations that facilitate tunnelling.
Equivalently, light scalar fields $h$ undergo fluctuations $\delta h \sim H/2\pi$ per $e$-folding.

We show that tunnelling from  de Sitter can be described by a simplified formalism assuming
that $H$ and the bounce energy density are sub-Planckian.
We also show how to include perturbatively the small extra Planck-suppressed corrections due to gravity.

Vacuum decay in the presence of gravity receives an extra contribution from another bounce, known as the
Hawking-Moss (HM) solution: a constant field configuration that sits at the top of the potential barrier~\cite{hawkingmoss}.
Being constant, it has the higher O(5) symmetry of 4-dimensional de Sitter space,
while the Coleman-de Luccia (CdL) bounce is only O(4) symmetric. The  Hawking-Moss contribution to  vacuum decay   vanishes for flat space, which corresponds to $H=0$.
We show that the Coleman bounce continuously deforms into the Hawking-Moss bounce,
and the two become equal at a critical value of $H$, usually equal to 1/2 of the curvature of the potential at its maximum.

Furthermore, the simplified formalism allows to easily include perturbatively the effect of a general non-minimal scalar coupling to gravity, as a modification of the effective scalar potential.

\medskip

The above features are relevant for the possible destabilization of the 
Standard Model (SM) vacuum during inflation.
The Higgs field can fluctuate towards
values a few orders of magnitude below the Planck scale, for which its potential
can be deeper than for the electroweak vacuum \cite{espinosariot,cosmo2,zurek,sibir,tetradis,Kohri:2016wof,rajantie,Enqvist,lalak}.
If vacuum decay happens during inflation, the regions of true vacuum expand and engulf the 
whole space~\cite{deluccia,tetradis,blau}. 
This catastrophic scenario is avoided
if $H$ is  small enough that vacuum decay is negligibly slow.
The derivation of a precise bound on the scale of inflation has been based 
mainly on a stochastic approach, relying on the numerical solution of 
Fokker-Planck or Langevin equations that describe the real-time evolution of
fluctuations in the scalar field.
We pursue here the alternative approach of computing the Euclidean tunnelling rate.
We reproduce some  previous results and correct others.


The paper is structured as follows.
In section~\ref{approx} we derive a simple approximation
for sub-Planckian vacuum decay. 
In section \ref{numeri} we validate the analytical expressions
by focusing on a toy renormalizable scalar potential,
and show how the Coleman bounce connects to the Hawking-Moss bounce.
In the final section~\ref{SM} we 
 obtain bounds on the scale of inflation $H$ by 
 computing the tunnelling rate, dominated by the Hawking-Moss bounce.
 Conclusions are given in section~\ref{concl}.

\section{General  theory and sub-Planckian approximation}\label{approx}
Coleman and de Luccia \cite{deluccia} developed the formalism for computing vacuum decay from a
de Sitter space with Hubble constant $H$ taking gravity into account.
In this section we review this formalism, extend it to a general non-minimal coupling of the Higgs to gravity and then derive simplified expressions that hold in the sub-Planckian limit $H, M \ll M_{\rm Pl}$,
where $M$ is the mass scale that characterises the scalar potential and thereby the bounce.
Within Einstein gravity this approximation applies to all cases of interest: indeed, Einstein gravity is non-renormalizable 
and must be replaced by some more fundamental theory at the Planck scale or below it.
Furthermore, the Hubble constant during inflation must be sub-Planckian to reproduce 
the smallness of the inflationary  tensor-to-scalar ratio.

\smallskip

Coleman and de Luccia assumed that the space-time probability density of vacuum decay is exponentially suppressed
by the action of a `bounce' configuration, like in flat space.\footnote{The proof valid in flat space cannot be extended
to de Sitter space because  the Hamiltonian is not well defined. 
One cannot isolate the ground state by computing a transition amplitude
in the limit of infinite time, because 
Euclidean de Sitter has  finite volume, so that 
multi-bounce configurations cannot be resummed in a dilute gas approximation.}
We consider the Euclidean action of a scalar field $h(x)$ in the presence of gravity,
\be
S=\int d^4x \sqrt{\det g} \left[\frac{1}{2}(\nabla h)^2+V(h) -\frac{\Rc}{16 \pi G}-
\frac{ f(h) }{2}\Rc \right]
\label{aaction} \ee
where $G\equiv 1/M_{\rm Pl}^2 \equiv 1/(8\pi\bp^2) \equiv \kappa/8\pi$ 
is the Newton constant, with  $\bar{M}_{\rm Pl}\approx 2.43\times 10^{18}\,{\rm GeV}$.
For the time being $V$ and $f$ are generic functions of the scalar field $h$. The action in eq.~(\ref{aaction}) is the most general action for the metric $g_{\mu\nu}$ and $h$ up to two-derivative terms: an extra generic function $Z(h)$ multiplying the kinetic term of the scalar field can be removed by redefining $h$. 
The classical equations of motion for gravity and for $h$ are~\cite{nucamendi}
\begin{eqnarray}
 (\bp^2+f)\left( \Rc_{\mu\nu}-\frac{g_{\mu\nu}}{2}\Rc \right)&=&
 \nabla_\mu h \, \nabla_\nu h -g_{\mu \nu}
\left[ \frac{(\nabla h)^2 }{2} + V  +\nabla^2 f\right] + \nabla_\mu \nabla_\nu f
\label{aeom1} \\
\nabla^2 h +\frac{1}{2} \frac{df(h)}{dh} \Rc  &=& \frac{dV(h)}{dh},
\label{aeom2}\end{eqnarray}
where $\nabla^2= \nabla_\mu  \, \nabla^\mu$.
The use of these equations allows us to simplify  the
action. Taking the trace of eq.~(\ref{aeom1}) one finds
\be
 \left({\bp^2}+f(h) \right)\Rc=(\nabla h)^2 +4 V(h)+
3 \, \nabla^2 f(h).
\label{Reom} \ee
Substitution in eq.~(\ref{aaction}) gives
\be
S=-\int d^4x \sqrt{\det g} \left[ V(h) +\frac{3}{2}  \nabla^2 f(h)  \right].
\label{aactioneom} \ee
The second term in the above expression is reduced to a boundary term upon
integration. For the problem at hand this term vanishes and one obtains 
\be S=-\int d^4x \sqrt{\det g} \, V(h).  \label{aactioneom2}\ee
 We are interested in the possible decay of the false vacuum during 
 a period in which the vacuum energy is dominated by a cosmological constant $V_0$.
We assume that  $V(h)$ has two minima, 
a false vacuum at 
$h=h_{\rm false}$ and the true vacuum at $h=h_{\rm true}$, 
with $h_{\rm true} >h_{\rm false}$. In the following we will set $h_{\rm false}=0$ without loss of generality.
The two minima are separated by a maximum of the potential
. 
We identify $V_0 = V(0)$ and split the potential as
 \be V(h)=V_0 + \delta V(h),\label{ParamPot} \ee such that $\delta V(0)=0$.  
 The vacuum energy density $V_0$ induces a de Sitter space with curvature 
$\Rc=12 H^2$, where the Hubble rate $H$ is given by
\be H^2 =\frac{ V_0}{3\bp^2
}\label{HLambda}\ee 
(we assume without loss of generality $f(0)=0$: a non-vanishing value of $f(0)$ can be absorbed in a redefinition of $\kappa$).
 In the following we will collectively denote with $\Phi_{\rm false}$ the field configuration with this de Sitter background and $h=0$. 
In the semiclassical (small $\hbar$)  limit the decay rate $\Gamma$ of the false vacuum  per unit of 
space-time volume ${\cal V}$ is given by  \cite{Coleman:1977py,Callan:1977pt,deluccia}
\be \frac{d\Gamma}{d\cal V} = A e^{-{\Sb}/\hbar} (1+{\cal O}(\hbar)), \label{GammaFormula}\ee
where $A$ is a quantity of order $M^4$, where $M$ is the mass scale in the potential.
The dominant effect is the bounce action ${\Sb}$.
 It  is given by  
\be {\Sb} = S(\Phi_B) - S(\Phi_{\rm false}),\label{Bdef}\ee where $\Phi_B$  is an unstable `bounce' solution of the Euclidean equations of motion,
such that  $\Sb$ is finite and there is no other configuration with the same properties and lower $\Sb$. In the rest of the paper we set the units such that $\hbar=1$.
In order to find $\Phi_B$ we follow~\cite{deluccia} and introduce an  O(4)-symmetric Euclidean ansatz for the Higgs field $h(r)$ and for the geometry
\beq  ds^2 = dr^2 + \rho(r)^2 d\Omega^2, \label{ansatz-metric} \eeq
where $d\Omega$ is the volume element of the unit 3-sphere.
On this background, the action becomes
\beq S = 2\pi^2 \int dr \rho^3\left[\left(
 \frac{h^{\prime 2}}{2}+V(h) \right)-
\frac{{\Rc}}{2\kappa}- \frac{{\Rc}}{2} f(h)\right], \label{Sexact}
\eeq
where the curvature is 
\be
{ \Rc}=-\frac{6}{\rho^3}(\rho^2 \rho''+\rho \rho^{\prime 2}-\rho)
\label{curv} \ee
and
a prime denotes $d/dr$. The simplified action of eq.~(\ref{aactioneom2}) becomes:
\be \bbox{S = -2\pi^2 \int dr\, \rho^3\, V(h)}.\label{ActionSimple}\ee
 The equations of motion   are
\begin{eqnarray}
h'' + 3\frac{\rho'}{\rho} h' &=& \frac{dV(h)}{dh} -\frac12 \frac{df(h)}{dh} {\Rc},
\label{eom1} \\
\rho^{\prime 2} &=& 1 + \frac{ \kappa \rho^2 }{3 (1+\kappa f(h))}\left(\frac{1}{2} h^{\prime 2}  -V(h)-3\frac{\rho'}{\rho} \frac{df(h)}{dh} h'\right).
\label{eom2} \end{eqnarray}
Let us discuss the boundary conditions. Since in the false vacuum  the space is a 4-sphere 
and topology cannot be changed dynamically,  the space described by $\rho$ will have the same topology; 
thus $\rho$ will have two zeros. 
One can be conventionally chosen to occur at $r=0$, and the other one at some value of $r$ that we call $r_{\rm max}$, 
\be \rho(0)=\rho(r_{\rm max})=0.\label{rhoBC}\ee
The whole space is covered by the
coordinate interval $[0,r_{\rm max}]$.
In the de Sitter case one has $r_{\rm max}=\pi/H$.
The equation of motion of $h$ in (\ref{eom1}) and  the regularity of $h$ at $r=0$ and $r=r_{\rm max}$ imply
 \be h'(0)=h'(r_{\rm max})=0.\label{BCh}\ee
In the limit of small $H$ (i.e.\ large $r_{\rm max}$),
the boundary condition $h'(r_{\rm max})=0$ implies $h(\infty) =0$
in view of the large volume outside  the core of the bounce.
Generically, in a non-trivial ($r$-dependent) bounce $h(r)$ does not tend to the false vacuum solution, $h_{\rm false}=0$, as $r\to r_{\rm max}$, unless we are in the flat space case, $r_{\rm max}\to \infty$.\footnote{This  can be shown whenever 
 $h$ and $\rho$ are regular functions at $r= r_{\rm max}$: 
 by Taylor-expanding  the equations of motion eq.s~(\ref{eom1})-(\ref{eom2})  around $r= r_{\rm max}$, using $h(r)= \sum_{n=1}^\infty c_n (r-r_{\rm rmax})^n$ and $\rho(r)=\sum_{n=1}^\infty a_n (r-r_{\rm max})^n$: one obtains a set of algebraic equations that force $c_n=0$.
Namely, for $r_{\rm max}<\infty$, the only regular function that goes indefinitely close to the false vacuum solution as $r\to r_{\rm max}$ is the false vacuum solution itself.
 }


 \subsection{The Hawking-Moss bounce}
The Hawking-Moss configuration \cite{hawkingmoss}, which we denote with   $\Phi_{\rm HM}$,  is a simple unstable finite-action solution satisfying the equations of motion and boundary conditions above. 
In this configuration the scalar sits at the constant value  $h=h_{\rm max}$ that maximizes\footnote{The
Coleman bounce is O(4)-symmetric.
The Hawking-Moss bounce, 
having a constant $h$, has the full O(5) symmetry of de Sitter space.
Vacuum decay at finite-temperature $T$ is described by a configuration with period $1/T$
in the Euclidean time coordinate.
At large $T$, the thermal bounce becomes constant in time, acquiring a O(3) $\otimes$ O(2) symmetry
(see~\cite{urbano} for a recent discussion).
These are different solutions:
a de Sitter space with Hubble constant $H$ 
is qualitatively similar but not fully equivalent to a thermal bath at temperature $T=H/2\pi$.}
\beq  \bbox{V_H(h) \equiv V(h)- 6 H(h_{\rm max})^2 f(h),}\label{eq:VHdef} \eeq
where $H(h)$ is the  Hubble constant given by 
\be H^2(h) = \frac{ \kappa V(h)}{3 (1+\kappa f(h))}. \label{HMH}\ee
The Hawking-Moss solution  exists whenever $V_H$ has a maximum.  
When $f=0$, $h_{\rm max}$ coincides with the maximum of the potential.
We can compute the tunnelling rate  by using the simplified action in eq.~(\ref{ActionSimple}), obtaining  for the Hawking-Moss solution 
\be {\Sb}_{\rm HM} =S(\Phi_{\rm HM})-S(\Phi_{\rm false})  = 24 \pi^2 \bp^4\bigg[\frac{1}{V_0} -\frac{ (1+\kappa f(h_{\rm max}))^2}{ V(h_{\rm max})}\bigg]. \ee
Defining $V_H (h) = V_0 + \delta V_H(h)$,
in the limit  
$\delta V(h_{\rm max})  \ll V_0$ and at leading order in $\kappa$  this formula simplifies to 
\be {\Sb}_{\rm HM}\simeq \frac{8\pi^2}{3}  \frac{\delta V_H(h_{\rm max})}{H^4},
\label{HMestim}\ee
where $H$ can be evaluated at $h=0$.
We shall examine the role of the Hawking-Moss solution for vacuum  decay, finding that
it is relevant for large values of $H$. 

\smallskip

Generically, there are also non-trivial solutions with a non-constant Higgs profile $h(r)$, which, in the flat-space limit,
reduce to the Coleman bounce~\cite{Coleman:1977py}. 
In order to determine the various bounces, one must solve the coupled eq.s~(\ref{eom1}) and (\ref{eom2}) with the boundary conditions described above.

 \subsection{Sub-Planckian approximation to the bounce}
The problem can be  simplified by using the low-energy approximation, which, as explained at the beginning of this section, is not physically restrictive if one works within the regime of validity of 
Einstein gravity (as we do). We illustrate now such an approximation.

 The low-energy approximation consists in assuming that gravity is weak in the sense that 
\be  H, \frac{1}{R}  \ll M_{\rm Pl}  \label{CondExp} \ee
where $R$ is the size of the bounce ($1/R$ is roughly given by the mass scale that appears in the 
scalar potential $V$).  
The two conditions arise because gravitational corrections are suppressed by powers of the Planck mass,
and are thereby small if the massive parameters of the problem are small in Planck units.
During inflation,  the condition on $H$ is satisfied in view of the experimental constraint $H<3.6\times10^{-5} \bp$ (see sec 5.1 of  \cite{Ade:2015lrj}).  The first condition is also not restrictive because it is necessary to avoid energies of order of the Planck scale, for which Einstein's theory breaks down.

Assuming that these conditions are satisfied, we expand $h$ and $\rho$ in powers of $\kappa = 1/\bp^2$:
\beq h(r) = h_0(r) + \kappa h_1(r)+{\cal O}(\kappa^2)  ,\qquad \rho(r) = \rho_0(r) + \kappa \rho_1(r)+{\cal O}(\kappa^2)\label{hrhoExp}.\eeq
The leading-order metric corresponds to de Sitter space:
\be \rho_0(r)=\frac{\sin(Hr)}{H}. \label{rho0dS} \ee 
Furthermore, we are interested in a situation in which $H \sim 1/R$:
otherwise one can neglect $H$ and return to the flat space approximation discussed in~\cite{tetradis,urbano}.
Thus we are in a regime in which the vacuum energy is dominated by $V_0  =3 H^2 \bp^2 \gg \delta V(h)$ 
and the gravitational  background is perturbed only slightly by the bounce $h(r)$.

The equation of motion of the zeroth order bounce $h_0(r)$ (that is the equation on the de Sitter non-dynamical background) and the first correction $\rho_1$ to the metric function can be obtained by inserting the expansion of eq.~(\ref{hrhoExp}) in eq.s~(\ref{eom1}) and (\ref{eom2}).
The de Sitter bounce $h_0(r)$ at zeroth order in $\kappa$ is given by
\be
\bbox{ h_0'' + 3 H \cot(H r)
 h_0' = \frac{dV_H(h_0)}{dh}  } \label{h0Eq}\ee
 where $V_H$ is given in eq.\meq{VHdef},
where $H$ can be evaluated at $h=0$ rather than at $h_{\rm max}$:
since the difference is an higher order effect in $\kappa$, we avoid introducing two different symbols.\footnote{As an aside comment,
an O(4)-symmetric space is conformally flat, such that by performing a Weyl transformation
one can revert to flat-space equations with a Weyl-transformed action.
A de Sitter space is conformally flat when written in terms of 
conformal time $\tilde r\equiv 2\tan (Hr/2)/H$.
Performing the associated Weyl transformation $h_0(r) = \tilde h_0(\tilde r)/ \cos^2(Hr/2)$, the bounce equation~(\ref{h0Eq})
becomes
\beq \label{eq:conformal}
\frac{d^2 \tilde h_0}{d\tilde r^2} + \frac{3}{\tilde r}\frac{d \tilde h_0}{d\tilde r} =\frac{\tilde V^{(1)}(\tilde h_0(1+ H^2\tilde r^2/4))}{(1+H^2 \tilde r^2/4)^3},\qquad
\tilde V \equiv V_H - H^2 h^2 = V -H^2(6 f+h^2) 
\eeq
where $\tilde V^{(n)}$ is the $n$-th derivative of $\tilde V$.}
The boundary conditions are
 \be h'(0)=h'(\pi/H)=0.\label{BCh0}\ee
At this lowest order, the effect of a general non-minimal coupling to gravity $f(h)$ is equivalent to replacing the potential $V(h)$ with the modified potential $V_H(h)$ given in eq.~(\ref{eq:VHdef}). 
The equation for $\rho_1$ is  
\be \left(\frac{\rho_1}{\cos(Hr)}\right)'  =  \frac{\tan^2Hr}{6H^2}\left(\frac{h_0'^2}{2} - \delta V(h_0)+3H^2f(h_0)-\frac{3H}{\tan Hr}\frac{df(h_0)}{dh}h_0'\right)\label{eq:rho1'}\ee
such that $\rho_1(r)$ can be obtained by solving 
 either by integration starting from $\rho_1(0)=0$
(although some care is needed to handle apparent singularities at $rH=\pi/2$),
or by converting eq.\meq{rho1'} into a 1st-order linear differential equation that can be solved numerically.
In the limit where the bounce has a size $R$ much smaller than $1/H$, the solution 
\be
\rho_1(r)\stackrel{R\ll 1/H}{\simeq} 
\cos (Hr)
\int_0^r dr \frac{r^2}{6}
\left( 
\frac{1}{2} \, h_0^{\prime 2} - \delta V(h_0)
-\frac{3}{r} \, \frac{df(h_0)}{dh} h'_0
\right)
\label{sol32} \ee
reduces to  the flat-space solution of~\cite{urbano}, times the overall 
$\cos(Hr)$ factor.

\medskip

Our goal now is to compute the action (difference) $\Sb$ of eq.~(\ref{Bdef}) because, which is the quantity that appears in the decay rate. The expansion for the fields in (\ref{hrhoExp}) leads to a corresponding expansion of $\Sb$ in powers of $\kappa$: 
\be {\Sb} = {\Sb}_0 + \kappa{\Sb}_1 + {\cal O}(\kappa^2).   \ee  
The zeroth order action is\footnote{The action contains the curvature term
enhanced by negative powers of the Planck mass. The ${\cal O}(1/\kappa)$ term cancels in  the difference defining $\Sb$, eq.~(\ref{Bdef}). 
Moreover, it leads to a term involving $\rho_1$ in the integrand of ${\Sb}_0$  proportional to  $(\sin(Hr)^2 \rho_1')'$.
However, this total-derivative term gives no contribution to ${\Sb}_0$ for a $\rho_1'$ that is regular at $r=0$ and $\pi/H$.} 
\be \bbox{ {\Sb}_0 =2\pi^2\int_0^{\pi/H} dr \frac{\sin^3Hr}{H^3} \left[\frac{h_0'^2}{2} +  \delta V_H(h_0) \right]}. \label{S0} \ee
The leading correction due to gravity, $\Delta {\Sb}_{\rm gravity}\equiv \kappa \Sb_1$, is
\bea \Delta {\Sb}_{\rm gravity}&=&
 \frac{6\pi^2}{\bp^2} \int_0^{\pi/H} dr \left[\frac{\sin^2(Hr)}{H^2} \rho_1 \left(\frac{h_0'^2}{2} +\delta V(h_0) - 3H^2 f(h_0)\right) - \frac{\sin(H r)}{H} \rho_1'^2 + \right. \nonumber \\
&& +2H\sin(H r)\rho_1^2  
 \left.  +\frac{\sin^2(Hr)}{H^2} f(h_0)\left(2H \cot(Hr) \rho_1'+\rho_1''\right) \right], \label{DSgrav}\eea 
The expression of $\Delta {\Sb}_{\rm gravity}$ above has been simplified by using the equation of $h_0$ in (\ref{h0Eq}) and by  an integration by parts.  
It can be further simplified as follows.
Rescaling $\rho_1(r) \to s \rho_1(r)$ corresponds to shifting $ \rho_1(r)$ by $(s-1) \rho_1(r)$. By noticing that $(s-1) \rho_1(r)$ is a particular variation $\delta \rho_1$ we conclude that the action 
must have an extremum at $s=1$. Applying this
argument to eq.~(\ref{DSgrav}) relates  the integrals of terms linear and quadratic in $\rho_1$.
The final simplified expression is:
 \beq \bbox{ \Delta {\Sb}_{\rm gravity}= \frac{6\pi^2}{\bp^2} \int_0^{\pi/H} dr \frac{\sin(H r)}{H} \left[ \rho_1'^2  -2 H^2 \rho_1^2  \right]}. \label{DSgrav2}\eeq
Note that the upper integration limit is simply $\pi/H$. In deriving it we have taken into account the dynamics of the spacetime volume: the shift in $r_{\rm max}$ does not affect the integral 
because the integrand contains a function, $\sin Hr$, which vanishes at the integration boundaries.\footnote{
The correction to $\rho$  generates a corresponding correction in $r_{\rm max}$, defined around eq.~(\ref{rhoBC}). Indeed
\be  0=\rho(r_{\rm max}) = \rho_0(r_{\rm max}) + \kappa \rho_1(r_{\rm max})+{\cal O}(\kappa^2) \label{rhoMax}\ee
tells us that $r_{\rm max}$ is a function of $\kappa$ that can be expanded around $\kappa=0$: 
$ r_{\rm max} = {\pi}/{H} + \kappa r_1+{\cal O}(\kappa^2)  \label{rmaxCorr} $.
By inserting the last expansion in eq.~(\ref{rhoMax}) we obtain 
$ \kappa \rho_0'(\pi/H) r_1 +\kappa \rho_1(\pi/H) +{\cal O}(\kappa^2) = 0$.
Noticing that $ \rho_0'(\pi/H) = -1$, we find 
$ r_1 = \rho_1(\pi/H), 
\label{r1}  $ 
where $\rho_1(\pi/H)$ can be obtained from the solution of eq.~(\ref{eq:rho1'}).}

In the limit $H\to 0$, eq.~(\ref{DSgrav2}) reduces to the flat-space expression found in~\cite{urbano}, which is
positive-definite, unlike the result for generic $H$.
Just like on flat space, $\Delta {\Sb}_{\rm gravity}$ is independent of $h_1$ on-shell: the reason is that the only way $h_1$ could appear at first-order in $\kappa$ is by taking the first variation of the $h$-dependent part of the action, eq.~(\ref{Sexact}), but this vanishes when $h_0$ solves eq.~(\ref{h0Eq}).

\medskip

In conclusion, eq.s~(\ref{h0Eq}), (\ref{eq:rho1'}), (\ref{S0}) and (\ref{DSgrav2}) tell us that, in order to compute the 
semiclassical decay rate including the first-order gravitational corrections,
one just needs to compute the 
bounce $h_0$ on the background de Sitter space. This is 
easier than solving the coupled equations for the bounce and the geometry in eq.s~(\ref{eom1}) and (\ref{eom2}). Being a one-dimensional problem, it can be solved through
an over-shooting/under-shooting method.
Then, one needs to plug $h_0$ in the expression for $\rho_1$ to get $\Sb_0+\Delta {\Sb}_{\rm gravity}$.
One can therefore  focus on the equation of $h_0$. 
Imposing the boundary conditions in eq.~(\ref{BCh0}) leads to well-defined solutions, as we will show  in the next sections.

\section{Renormalizable potential} \label{numeri}
In order to understand the influence of the de Sitter background on vacuum decay, we perform a numerical
study of the problem for a toy renormalizable  potential 
\be V(h)=V_0+\frac{M^2}{2} h^2-\frac{A}{3}  h^3 + \frac{\lambda}{4}  h^4 \label{eq:V4}\ee 
with $M^2 =\lambda  h_{\rm max} h_{\rm true}$ and $A = \lambda (h_{\rm max}+h_{\rm true})$,
such that the potential has a maximum at $h= h_{\rm max}$
and two vacua at $h=0$ and at $h=h_{\rm true}$:
the latter vacuum is the true deeper vacuum provided that $h_{\rm true}>2 h_{\rm max}>0$.
Quantum corrections are perturbatively small when $\lambda \ll 4\pi $ and $A\ll 4\pi M$.
The curvature of the potential at its maximum is $\mu^2 \equiv - V^{(2)}(h_{\rm max}) = \lambda (h_{\rm true}-h_{\rm max}) h_{\rm max}$.
The constant term $V_0$ gives a Hubble constant $H$ through eq.~(\ref{HLambda})
.


\begin{figure}[!t]
\centering
$$
\includegraphics[width=0.45\textwidth]{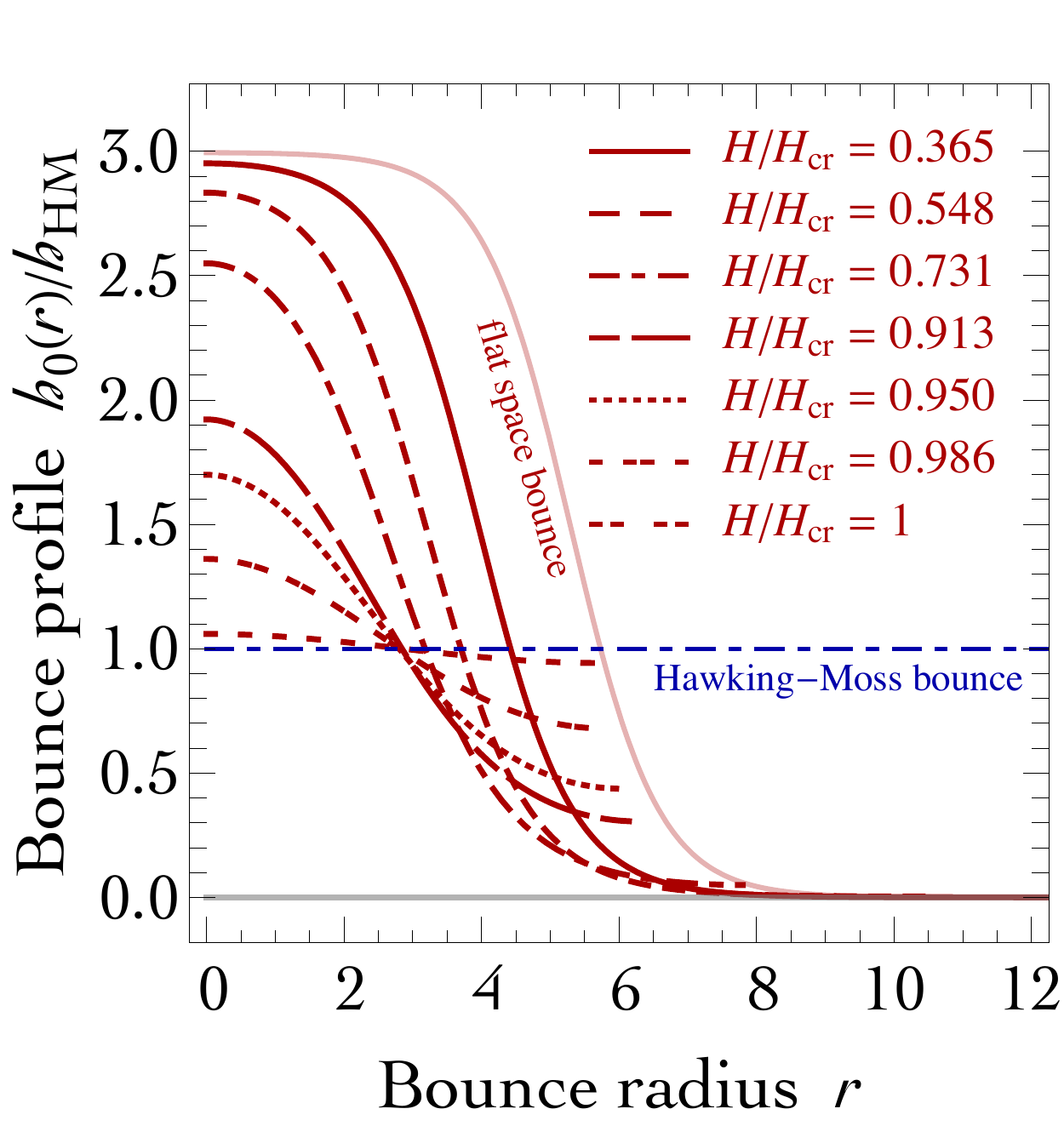}\qquad 
\includegraphics[width=0.45\textwidth]{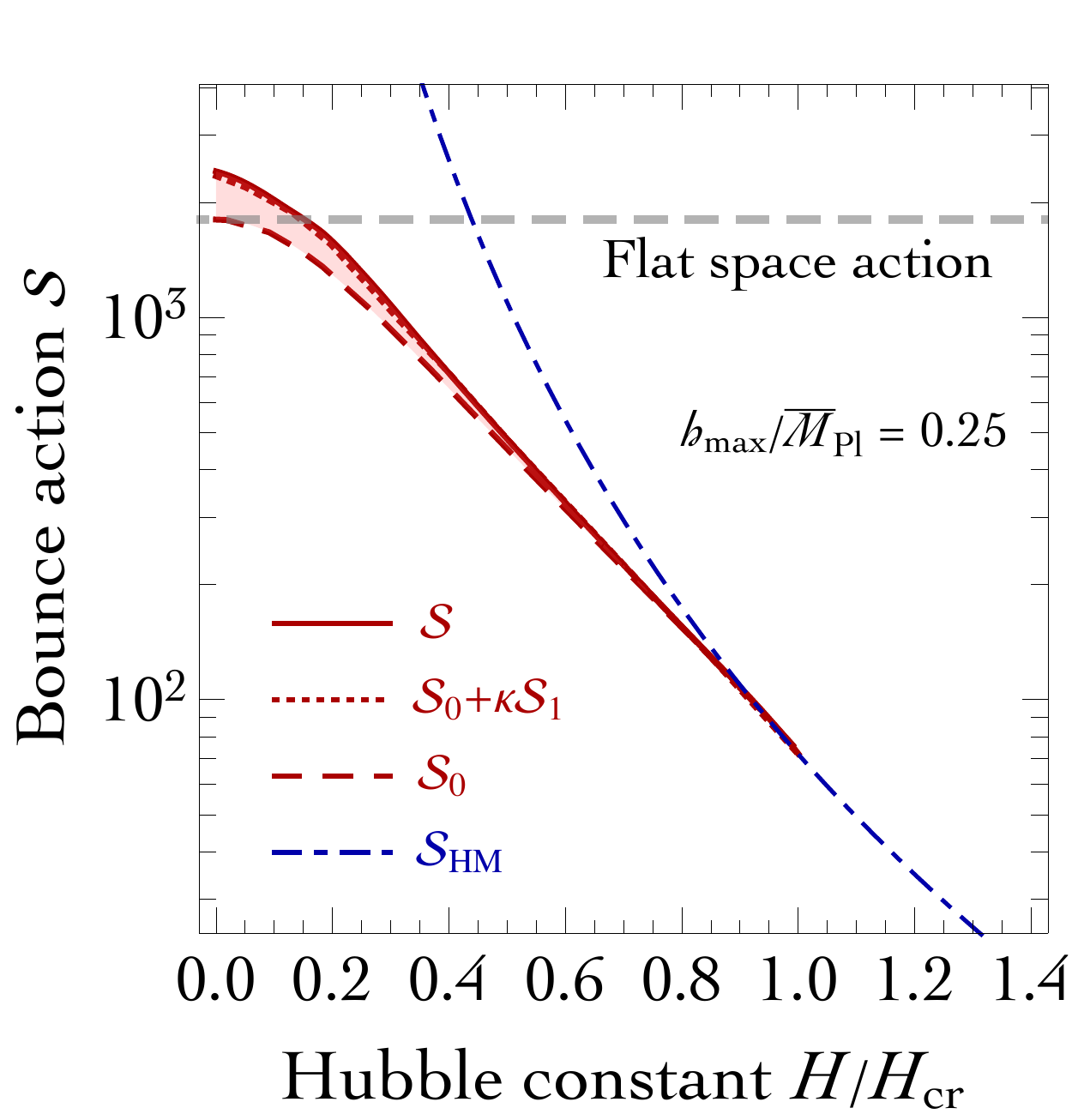}
 $$
\caption{\em\label{fig:0} 
We consider the renormalizable quartic potential of eq.\meq{V4} for quartic scalar coupling $\lambda=0.6$,
$h_{\rm true}=3h_{\rm max}$ and vanishing non-minimal coupling to gravity.
The results exhibit the typical features of the general case.
The left panel shows the bounce $h_0(r)$ for different values of the Hubble constant $H$.
As $H$ is  increased, the Coleman flat-space bounce gradually tends towards 
the Hawking-Moss bounce,  until
only the Hawking-Moss solution remains at $H>H_{\rm cr}$.
The right panel shows the Coleman and Hawking-Moss actions,
comparing our weak-gravity expansion with the full numerical result.
Fig.~\ref{fig:Sn} shows the actions of the extra multi-bounce solutions.
}

\end{figure}

\begin{figure}[!t]
\centering
$$
\includegraphics[width=0.65\textwidth]{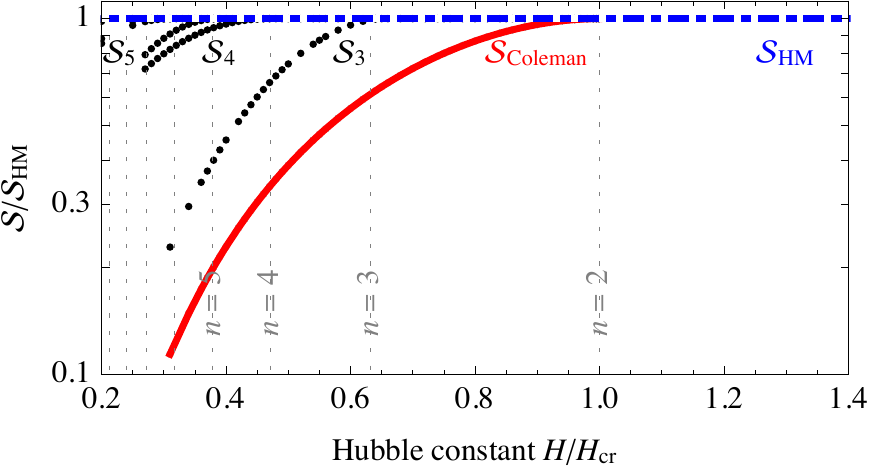}
 $$
\caption{\em\label{fig:Sn}  
For the  renormalizable quartic  potential considered in fig.~\ref{fig:0}, we show the
actions of the multi-bounce solutions in units of the Hawking-Moss action.}
\end{figure}

\subsection{Zeroth order in $H,M\ll M_{\rm Pl}$}\label{Renorm0}
The main qualitative influence of the Hubble rate $H$ on vacuum decay is most easily understood 
at zeroth order in the sub-Planckian expansion, 
ignoring the gravitational corrections that will be discussed in the next section.

We consider a typical illustrative example:
vanishing non-minimal coupling to gravity (introduced later),
$\lambda=0.6$ and
$h_{\rm true}=3h_{\rm max}$.
In fig.~\ref{fig:0} we show the resulting bounces $h_0(t)$ at zeroth order in $H,M\ll M_{\rm Pl}$
for increasing values of $H$.

For $H \ll M$ the de Sitter radius $1/H$ is much larger than the scale
of the  flat-space bounce, of order $1/M$.
Thereby, the flat space bounce is negligibly affected by the curvature of the space,
fitting comfortably into a horizon.

We see that the critical value above which $H$ starts
influencing the bounce action is of order $M$.
Thereby the bounce correction to the energy density is of order $M^4$,
which is negligible with respect to $V_0 =3 \bp^2 H^2$.
This confirms that, in the relevant range, the bounce correction to the background is negligible, being Planck suppressed,
so that it makes sense to first consider the zeroth-order approximation.

Fig.~\ref{fig:0}a shows that, increasing $H$, the flat-space bounce 
flattens and it tends to the constant Hawking-Moss bounce
$h_{\rm HM}(r) =h_{\rm max}$
above a finite critical value $H_{\rm cr}$ of $H$, of order $M$.\footnote{The thin-wall approximation
approximates the bounce as two different constants, at $r<R$ and $r>R$.
In cases where the thin-wall approximation holds in flat space at $H\ll M$, 
it ceases to be valid as $H$ is increased.
Thereby the continuous transition from the Coleman bounce to the Hawking-Moss bounce
is  not visible in the thin-wall approximation~\cite{Lee:2008hz,lalak}.
We emphasize that the Hawking-Moss bounce is not an approximation to the Coleman bounce: they are two different solutions.}

The critical value of $H$ can be analytically computed by approximating
 the potential as a quadratic Taylor series in $h$ around its maximum:
$V(h) \simeq V(h_{\rm max}) + \frac12  V^{(2)}(h_{\rm max}) (h- h_{\rm max})^2$, such that the bounce equation 
at lowest order in $\kappa$ becomes
\beq h''_0(r) + 3 H \cot(Hr) h'_0(r) \simeq  V^{(2)}(h_{\rm max}) [h_0(r) - h_{\rm max}]. \label{linEqH0}\eeq
This linear equation is solved by 
\beq h_0(r)- h_{\rm max} \propto  \frac{P_n^1(\cos (Hr))}{\sin (Hr)}\qquad\hbox{where}\qquad
n = \frac12 \bigg( \sqrt{9 - \frac{4V^{(2)}(h_{\rm max})}{H^2}} - 1\bigg)\eeq 
is the order
of the Legendre function $P_n^1$;
 the other independent solution is not regular in $r=0$.
The solution
diverges at 
$\pi/H$ unless  $n$ is integer.
For $n=1$ one gets the constant Hawking-Moss solution.
The first non constant solution, $h_0(r) - h_{\rm max} \propto \cos(Hr)$,
arises for $n=2$, corresponding to the critical value
\beq  H_{\rm cr}\equiv \sqrt{-V^{(2)}(h_{\rm max})}/2.\eeq
For values of $H$ close to $H_{\rm cr}$,
by expanding the potential to higher orders around its maximum,
one finds that  Coleman bounces must satisfy
\beq \label{eq:Delta}
\Delta \equiv 4 +\frac{V^{(2)}(h_{\rm max})}{H^2}  = - \frac{(h_0(0)- h_{\rm max})^2}{14H^2}\bigg[
V^{(4)}(h_{\rm max}) + \frac{V^{(3)}(h_{\rm max})^2}{12H^2} 
\bigg]
\eeq
and their action is
$\Sb_{\rm Coleman} \simeq \Sb_{\rm HM} + 2\pi^2 (h_0(0)- h_{\rm max})^2\Delta/15 H^2$~\cite{gr-qc/0409001}.
The sign of $\Delta$ is fixed by the potential, so that
Coleman bounces exist for $H<H_{\rm cr}$ when $\Delta<0$,
and for $H> H_{\rm cr}$ otherwise.\footnote{The expansion of the potential fails for
different potentials that involve vastly different mass scales (in particular the ones with a very flat barrier), which
need a more careful analysis of higher order terms in eq.~(\ref{linEqH0}).
In particular, if $V^{(2)}(h_{\rm max})$ vanishes, it gets replaced by an average around the top of the barrier~\cite{Jensen:1983ac,gr-qc/0409001,hep-th/0410142}.}
Our potential has $V^{(4)}>0$ and thereby $\Delta<0$, so that  Coleman bounces exist only for $H< H_{\rm cr}$.

\smallskip

Higher values of $n\ge 3$ correspond to bounces that cross the top of the potential $n-1$ times
and exist for $H \le \sqrt{-V^{(2)}(h_{\rm max})/(n^2+n-2)}$.
They never dominate the path-integral, as their action is between the Coleman action and the Hawking-Moss action,
as shown in fig.~\ref{fig:Sn}.
For $H \ll \sqrt{-V^{(2)}(h_{\rm max})}$ their actions, 
${\cal S}_n$, are multiple integers of the Coleman action,
allowing the resummation of their contributions in the dilute-gas approximation~\cite{Callan:1977pt}.
As $H$ grows, multi-bounce solutions progressively no longer fit into the Euclidean de Sitter space.

\subsection{First order in $H,M\ll M_{\rm Pl}$}

Fig.~\ref{fig:0}b shows the bounce actions.
In order to compute them, including gravitational corrections,
we need to fix the overall mass scale of the potential.
We choose a mildly sub-Planckian value of the field value of the top of the potential, $h_{\rm max} =0.25\bp$,
in order to have mild Planck-suppressed corrections.

\smallskip

The Hawking-Moss bounce exists whenever the potential has a barrier, and its action can be easily computed exactly.
When the Hubble constant is much smaller than the inverse size of the bounce, the Hawking-Moss bounce becomes
negligible because its action (plotted in blue) becomes large.

\smallskip

The Coleman bounce only exists for sub-critical values of $H$,
and its action (plotted in red) smoothly merges with the Hawking-Moss action for $H=H_{\rm cr}$.
We plotted 3 different results for its action:
\begin{enumerate}
\item The continuous curve is the full numerical result $\Sb$.
\item The dashed curve is $\Sb_0$, the action at zeroth order in $\kappa=1/\bp^2$.
We see that it already provides a reasonably accurate approximation.

\item The dotted curve  is
the first-order approximation $\Sb_0 + \kappa \Sb_1$,
and  it almost coincides with the exact numerical curve.

\end{enumerate}

\begin{figure}[t]
\centering
$$\includegraphics[width=0.45\textwidth]{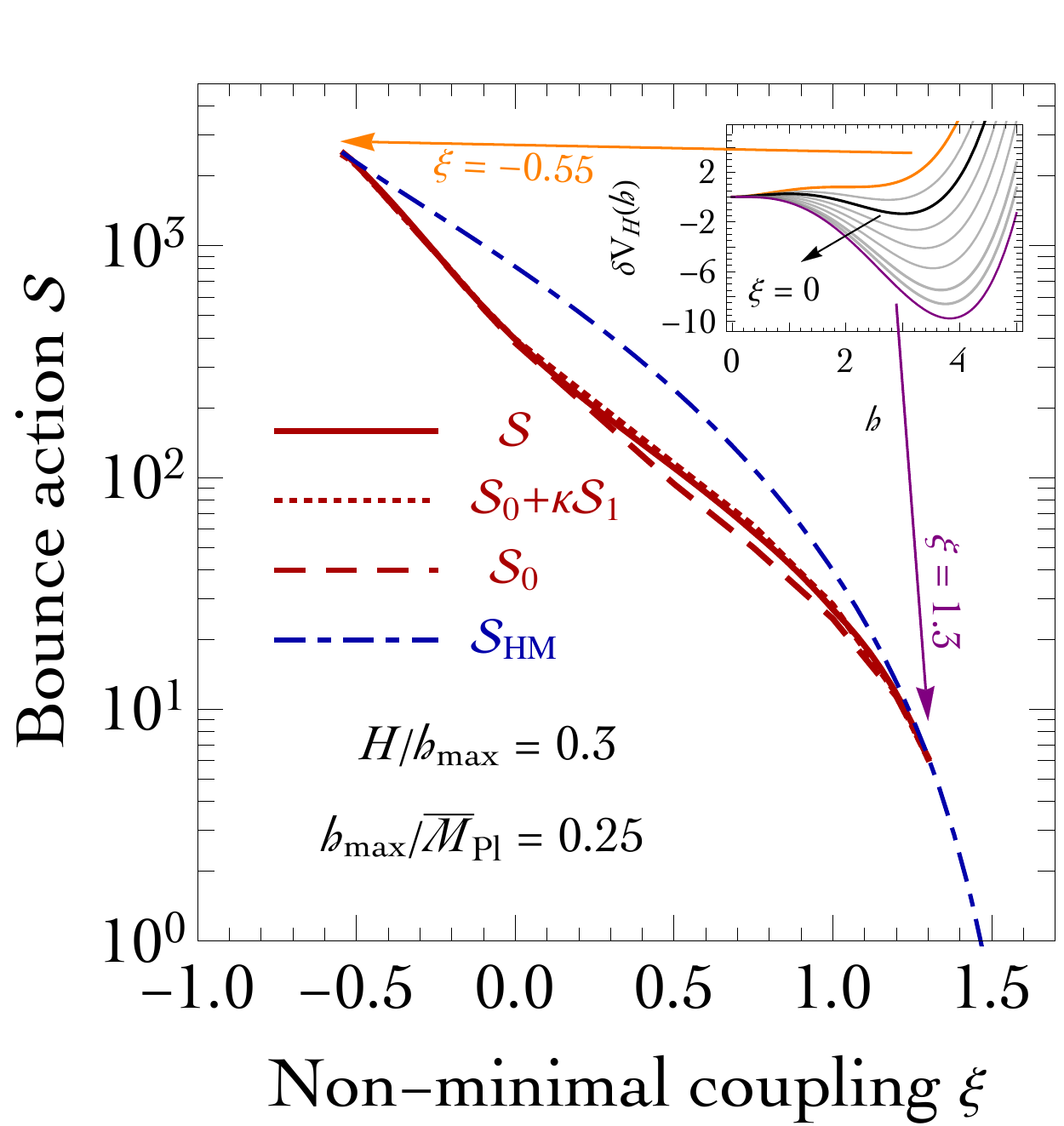} $$
\caption{\em\label{fig:CsiDependence} 
Coleman and Hawking-Moss bounce actions computed after adding a non-minimal coupling to gravity
$f = \xi h^2$ to the model considered in fig.~\ref{fig:0}.
Its main effect is to modify the effective potential, so that
the true vacuum disappears for negative large $\xi$, and the potential barrier disappears for positive large $\xi$.
Gravitational corrections to the action are again well approximated by our sub-Planckian expansion.}
\end{figure}

\subsection{Effect of non-minimal couplings}

Finally, the effect of a non-minimal coupling to gravity $f(h)$ can be trivially discussed
at zeroth order in the sub-Planckian limit,
being just  equivalent to a shift in the scalar potential
$V \to V_H =V- 6 H^2 f $, see eq.~(\ref{h0Eq}).
This effect is not present in flat space, where the  Hubble constant vanishes, $H=0$.

For example a non-minimal coupling to gravity $f(h)=\xi h^2$, 
governed by a dimension-less coupling $\xi$,
is equivalent to 
 a  shift $M^2 \to M^2_H \equiv  M^2 - 12\xi H^2$
 in our renormalizable quartic potential of eq.\meq{V4}.
This is just a  redefinition of the potential parameters
which makes the qualitative implications obvious, in agreement with the full numerical result shown in fig.~\ref{fig:CsiDependence}:
\begin{itemize}
\item 
A positive $\xi>0$ reduces $M^2_H < M^2$ and thereby the potential barrier,
decreasing the actions and increasing the tunnelling rate (see also~\cite{lalak}).
The critical value $H_{\rm cr}$ depends on $\xi$, so that
by increasing $\xi$ one first violates the condition $H < H_{\rm cr}$,
leading to the disappearance of Coleman bounces.
A larger $\xi$ leads to $M_H^2 <0$, so that the potential barrier disappears
and  the false vacuum is classically destabilised: this corresponds to $\Sb_{\rm HM}=0$ in fig.~\ref{fig:CsiDependence}.

\item A negative $\xi$ has the effect of increasing the potential barrier.
Ultimately, a too large negative $\xi$ destabilizes the true vacuum:
at this point $V(h_{\rm true}) = V(h_{\rm max})$, such that the
Coleman bounce becomes equal to the Hawking-Moss bounce.

\end{itemize}
In fig.~\ref{fig:CsiDependence} we also depict the full numerical action assuming a mildly sub-Planckian 
potential with $h_{\rm max} = 0.25\bar M_{\rm Pl}$: our sub-Planckian expansion again reproduces the full numerical result.
The above simplicity is lost if Planckian energies are involved; however in such a case Einstein gravity cannot be trusted.

\section{Standard Model vacuum decay during inflation}\label{SM}
Finally, we apply our general formalism to the  case of physical interest:
instability of the electro-weak SM vacuum during inflation.

\medskip

The probability per unit time and volume of vacuum decay during inflation
can be estimated on dimensional grounds as $d\wp/dVdt\sim H^4 \exp(-{\Sb})$,
where ${\Sb}$ is the action of the relevant bounce configuration.
The total probability of vacuum decay during inflation
then is $\wp\sim T L^3 H^4 \exp(-{\Sb})$, corresponding to a
total time $T\sim N/H$ and volume $L^3 \sim H^{-3} \exp (3N)$,
such that $\wp \sim \exp(3N - {\Sb})$.
Therefore, a small `probability' $\wp\sim 1$ of vacuum tunnelling during inflation
needs a bounce action
${\Sb}\gta 3N$.\footnote{Evading this bound trough anthropic selection
would need a very special landscape of unstable vacua with no low-scale inflation.}
The horizon of the visible universe corresponds to a
minimal number of $N \sim 60$ $e$-foldings of inflation.

\medskip

The computation of  the bounce actions ${\Sb}$ in the SM case needs to take into account the
peculiar features of the SM Higgs potential,
which is nearly scale-invariant  and can be approximated as
\be
 V (h)\approx \lx(h)\, \frac{h^4}{4}
\approx - b \ln \left(\frac{ h^2}{h_{\rm cr}^2 \sqrt{e}}\right) \frac{ h^4}{4}
\label{StrV}
\ee
where the running Higgs quartic can turn negative, $\lx (h)<0$,
at large field values.
This happens  for the present best-fit values of $M_t$, $M_h$ and $\alpha_3$,
that lead to $h_{\rm cr}=5\times 10^{10}\GeV$~\cite{1112.3022,1307.3536,tetradis}. 
The $\beta$-function of $\lambda$ around $h_{\rm cr}$ can be approximated as
$b \approx 0.15/(4\pi)^2$.

\smallskip

Furthermore, we add a  nonminimal coupling of the Higgs field to gravity  $f(h)=\xi_H h^2$, such that 
the effective potential of eq.\meq{VHdef} relevant in the sub-Planckian limit is
\beq V_H = V - 12 \xi_H H^2 \frac{h^2}{2}. \label{eq:VHSM}\eeq
Finally, we assume that inflation can be approximated as an extra constant term $V_0$ in the potential.
A possible extra quartic scalar coupling of the inflaton to the Higgs would manifest itself as
an extra  contribution to the effective Higgs mass term in eq.\meq{VHSM},
which is equivalent to a modified effective $\xi_H$.

\subsection{SM vacuum decay for small $H$}\label{SMsmallH}
In the limit of small $H$ one can view curvature, gravity and quantum effects as
perturbative corrections to the simple approximation of a constant $\lambda<0$.
Perturbing around a potential with no barrier and no true vacuum requires a careful understanding~\cite{Isidori:2001bm}.
Kinetic energy acts as a barrier, such that
the dimension-less potential admits a continuous family of flat-space bounces, parameterized by their arbitrary scale $R$:
\be \label{eq:h0}
h_0(r)= \sqrt{\frac{2}{|\lambda|}} \frac{2 R}{r^2+R^2}\qquad \hbox{($H=0$)}. \ee
The action of these `Fubini' bounces is $\Sb=8\pi^2/3|\lx|$~\cite{rychkov}. Minimal gravitational couplings have been included numerically in \cite{Lee:2012ug,Lee:2014ula}; in this section we will apply our analytic sub-Planckian approximation to take into account gravity (including non-minimal couplings).

Quantum corrections can be included roughly by renormalizing the quartic coupling at the scale of the bounce,
$\Sb \approx 8\pi^2/3|\lx(1/R)|$~\cite{Isidori:2001bm}.
For the best-fit values of the SM parameters, the Higgs quartic $\lambda$ runs in such a way
that tunnelling is dominated by mildly sub-Planckian bounces,
such that  Planck-suppressed corrections are small in flat space~\cite{rychkov,urbano}.

We now include curvature and gravitational effects, assuming $H \ll 1/R$.
Performing the integral in eq.~(\ref{sol32}) we obtain
the leading correction in small $H R$ to the metric:
\be
\rho_1(r)= \cos(H r)\frac{1 + 6 \xi_H}{3|\lx|R} \left( \frac{r R (r^2 - R^2)}{(r^2 + R^2)^2} + \arctan \left( \frac{r}{R} \right) \right) .
\label{rho1int} \ee
The gravitational corrections to the action
combine with the quantum corrections (see also~\cite{sibir}) 
in order to give the final formula
valid for $H \ll 1/R \ll M_{\rm Pl}$:
\be 
\label{eq:SSM}
 {\Sb} \simeq \min_{R\ll 1/H} \left\{ \frac{8\pi^2}{3|\lambda (1/R)|}
 \left[1+6(1+6\xi_H)(H R)^2 \ln  H R \right]
 +\frac{32\pi^2 (1+6\xi_H)^2}{45 (R \bar{M}_{\rm Pl} \,\lambda(1/R))^2}\right\}.
\ee
This expression only holds when the corrections are small. In this regime, the
vacuum decay rate during inflation is negligible, having a rate similar to the 
rate in the longer post-inflationary phase.

\subsection{SM vacuum decay for large $H \gg  h_{\rm cr}$}\label{toySM}
The interesting case that can lead to possibly significant inflationary
 enhancements of the tunnelling rate corresponds to values
of $H$ comparable or larger than the inverse size $1/R$ of the bounce, 
so that the approximation used in the previous section~\ref{SMsmallH} does not apply.

For simplicity, we start by discussing the opposite limit, in which  
the Hubble constant $H$ during inflation is sub-Planckian and much larger than the critical
scale $h_{\rm cr}$ above which the Higgs quartic coupling turns negative.
Then,  we can approximate
the SM potential at field values around $h \sim H$ as a quartic potential, 
with a constant negative $\lambda$ renormalized around $H$.
Adding the non-minimal coupling to gravity $f(h)=\xi_H h^2$ gives the
effective potential relevant in the sub-Planckian limit:
\beq \label{eq:SMlargeH}
\delta V_H = - 12 \xi_H H^2 \frac{h^2}{2} + \lambda \frac{h^4}{4}.\eeq
Scale invariance is broken by the de Sitter background with Hubble constant $H$,
so that 
the bounce action  can only depend on the dimensionless parameters $\xi_H$ and $\lambda$.
By performing the field redefinition $h(r) \to \alpha h(r)$, where $\alpha$ is a constant, one obtains 
$\Sb(\xi_H,\lambda) = \alpha^2 \Sb(\xi_H, \alpha^2\lambda)$ 
which implies  ${\cal S}\propto 1/\lambda$.  
Therefore, we only need to compute $\Sb$ as function of $\xi_H$.

A potential barrier  exists for $\xi_H <0$: then $h_{\rm max} = H \sqrt{12\xi_H/\lambda}$ and 
Hawking-Moss bounces have action $\Sb_{\rm HM}=-96\pi^2\xi_H^2/\lambda$.

According to the argument in section~\ref{Renorm0},
the critical value of $H$ that controls the existence of Coleman bounces
is $H_{\rm cr} = \sqrt{-V^{(2)}(h_{\rm max})}/2$.
For the potential of eq.\meq{SMlargeH} this means
$H/H_{\rm cr} = 1/\sqrt{-6\xi_H}$, so that $H = H_{\rm cr}$ for
$\xi_H = -1/6$.
As discussed in section~\ref{Renorm0}, Coleman bounces exist for $H < H_{\rm cr}$ or $H > H_{\rm cr}$ depending
on the sign of the higher-order coefficient $\Delta$ defined in eq.\meq{Delta}.
In the present case $\Delta \propto -\lambda(1+6\xi_H)$, with a positive proportionality constant.
This potential behaves in an unusual way: $\Delta$ vanishes at the critical value $\xi_H=-1/6$,
for which Hawking-Moss bounces have the same action ${\cal S}=-8\pi^2/3\lambda$ as flat-space Fubini bounces.
Flat-space bounces are relevant because the action is Weyl invariant for  $\xi_H=-1/6$,
such that de Sitter space is conformally equivalent to flat space.
Indeed, the Weyl-transformed eq.\meq{conformal} 
reduces to $\tilde h''_0 + 3\tilde h'_0/\tilde r = \lambda \tilde h_0^3$, satisfied
by  Fubini bounces
$\tilde h_0 = \sqrt{2/|\lambda|} 2R/(R^2+\tilde r^2)$, where $R$ is an arbitrary constant.
By Weyl-rescaling them back to the original field $h_0$ and coordinate $r$ we obtain 
the Coleman bounce  for $\xi_H =-1/6$, and thereby $H = H_{\rm cr}$:
\beq \label{eq:h0conf}
h_0(r) = \frac{\sqrt{2/|\lambda|} H^2R}{1 +H^2 R^2/4 - (1-H^2R^2/4)\cos Hr }.\eeq
For $R=2/H$ this corresponds to constant Hawking-Moss bounces; for $R\ll 2/H$ to Coleman bounces,
for $R\gg 2/H$ to Coleman bounces centred around $r = \pi/H$.
For this special potential 
the convergence of Coleman bounces with Hawking-Moss bounces happens at $H = H_{\rm cr}$
rather than gradually for $H \to H_{\rm cr}$.

\begin{figure}[t]
\centering
$$
\includegraphics[width=0.5\textwidth]{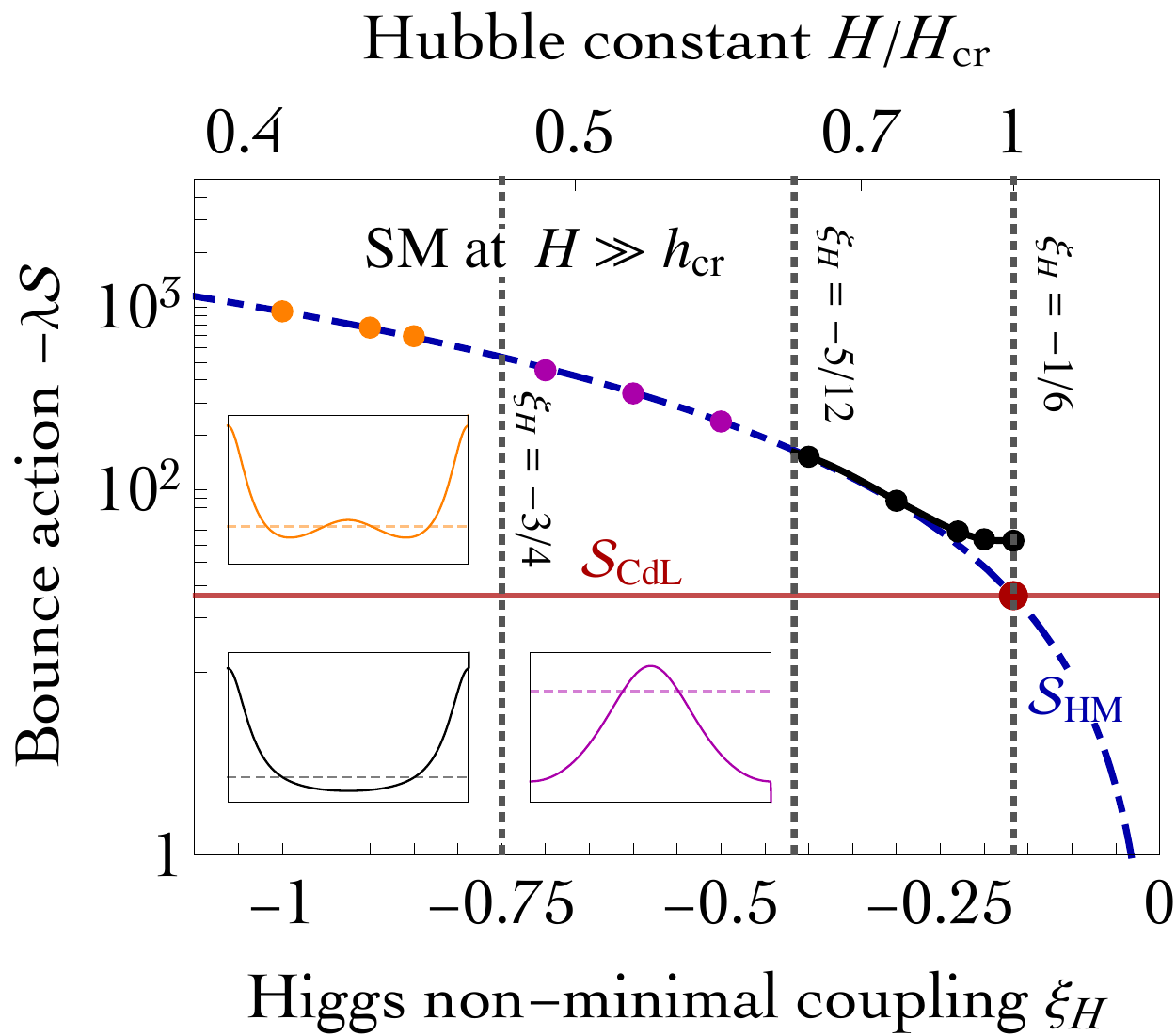}
 $$
\caption{\em\label{fig:toySM} 
Bounce actions for the potential
$V=\lambda \sfrac{h^4}{4}$ with $\lambda<0$ and a non-minimal $\xi_H$ coupling to gravity.
As the action is proportional to $1/\lambda$, we plot the product $-\lambda\Sb$ as function of $\xi_H$.
The red dot denotes the Coleman bounces of eq.\meq{h0conf}.
The other dots denote multi-bounces, as plotted in the insets.}
\end{figure}

Eq.~(\ref{eq:h0conf}) together with de Sitter space
actually solve the full gravitational equations (not only those in the sub-Planckian limit), 
as can be explicitly verified or understood from the argument given in flat space in footnote 6 of~\cite{urbano}.

\medskip

Fig.~\ref{fig:toySM} illustrates the situation.
Extra multi-bounce solutions that perform multiple
oscillations around the maximum with action ${\cal S} \ge {\cal S}_{\rm HM}$
appear below the extra critical values $\xi_H =(2-n-n^2)/24$ with $n=\{3,4,\ldots\}$.


\medskip

In conclusion, vacuum decay is safely slow if $\Sb_{\rm HM}>3N$ which implies the condition
\beq\label{eq:bigH} -\xi_H<\sqrt{-\frac{N\lambda(H)}{32\pi^2}} \approx -0.04 \sqrt{-\frac{\lambda(H)}{0.01}}\qquad
\hbox{for $H\gg h_{\rm cr}$}. \eeq

%
%
%
%
%
%
%
%

\subsection{SM vacuum decay for $H\sim h_{\rm cr}$}
We finally consider the more complicated intermediate case where the Hubble constant $H$ during inflation
is comparable to the instability scale $h_{\rm cr}$ of the SM potential.
For best-fit values of the SM parameters, this scale is sub-Planckian, so that
we can compute at zeroth order in $1/\bp$ by solving the approximated bounce equation, eq.~(\ref{h0Eq}).


\subsubsection*{SM vacuum decay for $\xi_H=0$}
Let us first consider the case of vanishing non-minimal coupling of the Higgs to gravity.
The approximated SM potential of eq.~(\ref{StrV})
has a maximum at $h_{\rm max}=h_{\rm cr}$.
The action of the Hawking-Moss bounce is
${\Sb}_{\rm HM}  = b\pi^2 h_{\rm max}^4/3H^4$ and
the critical value of $H$ is
\beq H_{\rm cr} \equiv \frac{\sqrt{-V^{(2)}(h_{\rm max})}}{2}  = \sqrt{\frac{b}{2} } h_{\rm max} =0.025 h_{\rm max} .
\eeq
At the critical value $H=H_{\rm cr}$ 
where Coleman and Hawking-Moss solutions merge, 
vacuum decay is suppressed by large actions ${\Sb}_{\rm HM} = 4\pi^2/3b \approx 13000$,
and the coefficient $\Delta$ that determines the behaviour of Coleman solutions 
is positive, such that Coleman bounces exist for $H > H_{\rm cr}$ and are irrelevant in view of 
$\Sb_{\rm Coleman} > \Sb_{\rm HM}$.
Thereby, the bound on vacuum decay is dominated by 
Hawking-Moss bounces at  $H > H_{\rm cr}$.
Imposing ${\Sb}_{\rm HM} \lta 3N$ implies
\be
\frac{H}{\hmax}\lta \left(\frac{8\pi^2}{9N} 
\frac{ V(h_{\rm max})}{\hmax^4} \right)^{1/4}  = 
\left(\frac{b\pi^2}{9N} \right)^{1/4} 
 \lta 0.06.
\label{bound1} \ee
This bound can be compared with the
bound $H/\hmax \lta 0.045$ derived in~\cite{tetradis} by solving the Fokker-Planck or Langevin equation
that describes the evolution in real time of the inflationary  quantum fluctuations of the Higgs field.
There is numerical agreement, even though the parametric dependence
in eq.~(\ref{bound1}) does not match the one in~\cite{tetradis}.
Indeed~\cite{tetradis} found that the Higgs acquires a Gaussian distribution with variance that grows
with the number $N$ of $e$-foldings as
$\sqrt{ \langle  h^2\rangle} = H \sqrt{N}/{2\pi} $,
without approaching a limiting distribution.
Thereby the dedicated study of~\cite{tetradis} was necessary and cannot
be reproduced by the Hawking-Moss  tunnelling computations,
which anyhow give a correct result, up to factors of order $\sqrt{N}$.

\begin{figure}[t]
$$\includegraphics[width=0.75\textwidth]{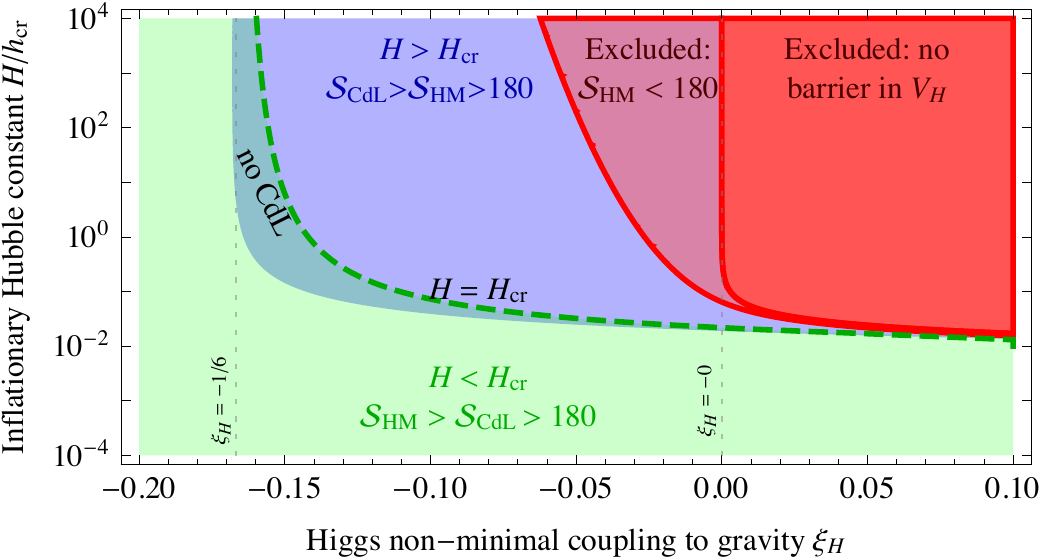}$$
\caption{\em  As a function of $\xi_H$ and of the Hubble constant in units of the instability scale $h_{\rm cr}$ 
(and for $N=60$ $e$-folds of inflation), we plot:
 the allowed region where $H < H_{\rm cr}$ in green with dashed boundary;
the region where tunnelling is dominated by Hawking-Moss in blue;
in red the excluded region where  tunnelling is too fast.
For $\xi_H<0$ this regions agrees with~\cite{tetradis}, while for $\xi_H>0$ our bounds are weaker.
\label{RHboundsNew}}
\end{figure}

\subsubsection*{SM vacuum decay for  $\xi_H\neq 0$}
A non-minimal coupling of the Higgs to gravity, $\xi_H\neq 0$, is unavoidably generated by SM RGE running, and
gives an extra contribution $M_h$ to the effective Higgs mass parameter
in the effective inflationary potential $V_H$ of eq.\meq{VHSM}, such that its maximum gets shifted from $h_{\rm max}=h_{\rm cr}$
to
\beq h_{\rm max} =  H \bigg[- \frac{b}{12\xi_H} W \bigg(-\frac{12\xi_H H^2}{b h_{\rm cr}^2}\bigg)\bigg]^{-1/2}
\eeq 
where $W(z)$ is the product-log function defined by $z=We^W$:
it is real for $z>-1/e$.
Otherwise, the potential $V_H$ has no  potential barrier because $M_h^2$ is too negative:
\beq  M_h^2 = -12\xi_H H^2< -\frac{b h_{\rm max}^2}{e} \label{eq:hmaxHSM}\qquad\hbox{(no barrier in $V_H$)}.
\eeq 
The action of the Hawking-Moss bounce is 
\be
{\Sb}_{\rm HM} 
\simeq \frac{8 \pi^2 }{3} \frac{\delta V_H(h_{\rm max})}{H^4}=\frac{48\pi^2 \xi_H^2}{b}\frac{1+2W}{W^2}.
\label{HMxi} \ee
The conditions that determine the relative role of Coleman and Hawking-Moss bounces around
the critical situation $H=H_{\rm cr}$ give rise to a non-trivial pattern, plotted in  fig.~\ref{RHboundsNew}.
The conclusion is again that only relevant vacuum decay bound is ${\Sb}_{\rm HM} > 3N$.
In the limit $H\gg h_{\rm cr}$ this reduces to eq.\meq{bigH}.
For generic values of $H$ the bound is plotted numerically in fig.~\ref{RHboundsNew}, where the red region is excluded
because inflationary vacuum decay is too fast.
For $\xi_H <0$  such bound agrees with the corresponding result of~\cite{tetradis}. 
Indeed~\cite{tetradis} found that a positive $M_h^2>0$
limits the Higgs fluctuations which, after a few $e$-foldings, converge towards a limiting distribution,
well described by the Hawking-Moss transition.
On the other hand, for $M_h^2\le 0$ ($\xi_H\ge 0$) the Higgs fluctuations
grow with $N$ 
(as $\sqrt{N}$ for $M_h^2=0$, and exponentially for $M_h^2<0$),
so that the detailed dynamical study of~\cite{tetradis} is needed.
Nevertheless, for $\xi_H>0$, the bound ${\Sb}_{\rm HM} > 3N$, is almost numerically
equivalent to the simpler bound on $M_h^2$ in eq.\meq{hmaxHSM}
that guarantees that $V_H$ has a potential barrier, which reads 
\be
\frac{H}{\hmax}\lta\left(\frac{b}{12e\xi_H}  \right)^{1/2}  \lta \frac{0.005}{\sqrt{\xi_H}}.
\label{bound3} \ee
Fig.~\ref{RHboundsNew} shows that this bound is weaker than the bound of~\cite{tetradis}.
A Langevin simulation performed along the lines of~\cite{tetradis} agrees with  our bound,
while~\cite{tetradis} made a simplifying approximation (`neglecting the small Higgs quartic coupling')
which is not accurate around the bound at $\xi_H>0$.


\section{Conclusions}\label{concl}
In section~\ref{approx} we developed a simplified formalism for computing
 vacuum decay from a de Sitter space with Hubble constant $H$,
assuming that both $H$ and the mass scale of the scalar potential are sub-Planckian.
This is not a limitation, given that otherwise Einstein gravity cannot anyhow be trusted. 
In this approximation, the bounce action is obtained as a power series in $1/M_{\rm Pl}$,
and a non-minimal scalar coupling to gravity
can be reabsorbed in an effective scalar potential, see eq.\meq{VHdef}.

In section~\ref{numeri} we considered a renormalizable single-field potential.
We verified that our expansion reproduces full numerical result.
Furthermore we found that, increasing $H$, the flat-space Coleman bounce 
continuously deforms into the Hawking-Moss bounce.
Only the Hawking-Moss bounce exists above a critical value of the Hubble constant,  equal to $H^2_{\rm cr}=-V^{(2)}(h_{\rm max})/4$.
For $H< H_{\rm cr}$ the Coleman bounce appears and
dominates vacuum decay, having a smaller action than the Hawking-Moss bounce.
For smaller values of $H$ extra bounces that oscillate around the top of the barrier appear,
but they never dominate the path-integral.
In the flat space limit $H\to 0$ they reduce to the infinite series of multi-bounce solutions.

\smallskip

In section~\ref{SM} we studied quantum tunnelling of the electroweak vacuum during inflation, 
assuming that the SM Higgs potential is unstable at large field values,
as happens for present central values of the SM parameters. 
Coleman bounces are still connected to Hawking-Moss bounces,
altough the fact that the SM potential has a negative quartic at large field values
and is quasi-scale-invariant makes their relation different.
We exhibit a limit where they are conformally equivalent.
Anyhow we found that only Hawking-Moss bounces imply a significant
bound on vacuum decay during inflation.
If the minimal coupling of the Higgs to gravity is negative, $\xi_H <0$, 
our  tunnelling computation confirms previous upper bounds on $H$ obtained from
statistical simulations (needed to address other cosmological issues).
If $\xi_H>0$ we find weaker bounds, and explain why the approximation made in earlier works~\cite{tetradis}
is not accurate.


\small

\subsubsection*{Acknowledgments}
We  thank G. D'Amico, J. R. Espinosa, A. Rajantie, M. Schwartz and S.M. Sibiryakov
 for useful discussions.
This work was supported by the ERC grant NEO-NAT.
The work of A.\ Katsis is
co-financed by the European Union (European Social Fund - ESF) and 
Greek national funds through the action ``Strengthening Human Resources 
Research Potential via Doctorate Research" of State Scholarships 
Foundation (IKY), in the framework of the Operational Programme ``Human 
Resources Development Program, Education and Lifelong Learning" of 
the National Strategic Reference Framework (NSRF) 2014 -- 2020.

    \end{document}